\documentstyle[12pt,epsfig]{article} 
\begin{document} 
\begin{center} 
\vspace{.2in} 
{\LARGE WHITHER NUCLEAR PHYSICS ?} 
\end{center} 
\vspace{.4in} 
\begin{center} 
{\bf Afsar Abbas}\\ 
\vspace{.05in} 
Centre for Theoretical Physics\\ 
JMI, Jamia Nagar, New Delhi-110025, India\\ 
\vspace{.05in} 
email afsar.ctp@jmi.ac.in 
\end{center} 
\vspace{.5in} 
\begin{center} 
{\bf Abstract} 
\end{center} 
\vspace{.3in}

Nuclear Physics has had its ups and downs. However in recent years,
bucked up by some new and often puzzling data, it has become a
potentially very rich field. We review some of these exciting developments
in a few important sectors of nuclear physics.
Emphasis shall be on the study of exotic nuclei and the new physics
that these nuclei are teaching us.

\newpage

Nuclear physics, perhaps not unlike any other discipline of 
scientific enterprise, has had mixed fortunes. 
Having had its heydays in the mid-twentieth century, 
nuclear physics' 
fortunes had sunk rather low towards the last decades 
of the 20th century. In fact, nuclear physics 
had been pronounced dead by many an expert. 
But as in Mark Twain's case, that announcement was rather 
"premature". Being a human, Mark Twain had to pass away ultimately, 
but in the case of nuclear physics, in recent years, there has been a 
great upsurge of scientific interest and at present 
nuclear physics is well and kicking.
The reasons which have led to this renaissance in nuclear physics, 
shall be the focus of this paper.

Most of the knowledge of nuclear physics, until recently, was based on 
radioactive decay studies 
and by nuclear reactions induced by beams of some 283 
species of stable or long lived nuclei one finds on earth. So to say, 
this has resulted in what may be called the conventional nuclear physics.
However, this allows us to study only somewhat limited regimes of nuclear 
physics. The nucleus, consisting of A number of nucleons, is governed by 
a large number of degrees of freedom. 
We have to judiciously choose the ones which are relevant; 
relevant, both in terms of theoretical challenges as well as  
in terms of the kind of experiments we are capable 
of performing today and as to which nuclear degrees of freedom do these 
experiments allow us to explore. Given what is being actively pursed 
today, we may define a three dimensional landscape. The three parametric 
space within which we may restrict our discussion are: 
firstly the temperature or excitation energy, 
secondly the angular momentum space and thirdly the space of the 
neutron-proton ratio ${(N-Z)} \over A$ [1].

The first two of the above can be studied by varying the combination of 
the target and the projectile and by changing the energy of the 
projectiles. There has been great
progress in studying nuclei at higher energies. As we have gone to higher 
and higher energies, we continue to improve our knowledge and 
understanding of nuclear physics. 
The frontier area in this parameter space today is 
the ongoing attempts to achieve the new phase of quark gluon plasma in 
heavy ion collisions at laboratories like CERN and RHIC [2,3,4].

Very high angular momenta of 50-80 has been achieved by grazing angle 
collisions between heavy ions. Hence motion of individual nucleons under 
these special conditions provide new insights into the nuclear 
dynamics [5]. The search for superheavy nuclei comes under the 
same category.

The studies from the first two parameter space have been very exciting 
in nuclear physics and have resulted in major improvements in our 
understanding of the nuclear phenomenon. However the most significant 
developments which are providing the intriguing possibility of 
some 'new physics" in  the regime of nuclear physics, have come from the 
developments arising from the third parametric space, ie those
brought in by changing the neutron -proton ratio in nuclei. 
And we shall concentrate upon these here. 

Drip line nuclei are those wherein the last neutron or proton is barely 
bound. As compared to the some 283 stable nuclei, the number of nuclei 
between the neutron and proton drip lines is close to 7000. This allows for
tremendous scope of variation of both the number of protons and neutrons.
The recent exciting development of the
generation of radioactive beams has made the  
above studies possible [6]. 
Study of exotic nuclei through these 
radioactive beams is transforming the whole landscape of nuclear 
physics, and as perhaps even more significantly,
is forcing us to attain and 
acquire a better understanding on what nuclear physics is all about. 

The most significant discovery in the sphere of exotic nuclei is 
certainly that of the neutron halo nuclei [7,8].
In conventional nuclei the rms radius of a nucleus is given by

\begin{equation}
R = {R_0} A^{1 \over 3}
\end{equation}

where ${R_0}$ is $\sim 1.2 \times {10}^{-13}$ cm.
In halo nuclei, it was found that the size is significantly larger than 
what the above formula gives and 
which holds well for conventional nuclei.
So for example for ${^{11}_{3} Li_{8}}$, the radius was found to be about 
3.3 fm which is much larger than what the above formula would give and is 
close to the radius of a nucleus with A=48 or so. In contrast the radius 
of ${^{9}_{3} Li_{6}}$ is only about 2.3 fm.  In the case of
${^{6}_{2} He_{4}}$ the rms radius determined empirically was 2.57 fm 
while that of ${^{4}_{2} He_{2}}$ is only 1.46 fm.
The two neutron separation energy of ${^{11}_{3} Li_{8}}$ and 
${^{6}_{2} He_{4}}$ are of very small value 0.3 Mev and 1.0 Mev 
respectively. Hence one assumes that these large halo nuclei 
${^{11}_{3} Li_{8}}$ and ${^{6}_{2} He_{4}}$
consist of a compact core
( of ${^{9}_{3} Li_{6}}$  and  ${^{4}_{2} He_{2}}$ respectively ) 
and with two extra neutrons orbiting around the core at a large 
distance away from it [7,8].

Hence the two extra neutrons in the above Li and He nuclei are very weakly 
bound and thus extend to large distances in these so called halo nuclei.
In addition it turns out the addition of the two extra neutrons outside
${^{9}_{3} Li_{6}}$ and outside ${^{6}_{2} He_{4}}$ cores, practically do 
not modify the electro-magnetic properties of the cores in 
${^{11}_{3} Li_{8}}$  and ${^{6}_{2} He_{4}}$.
For example the magnetic dipole moment of $^{9} Li$ and $^{11} Li$
are 3.4 and 3.7 nm respectively while the two electric quadrupole moments 
are -27.4 and -31 mb respectively. 
Charge changing reaction for $^{8,9,11} Li$ on carbon target
are nearly the same for all the Li isotopes [8]. This shows that the 
charge distribution in all these nuclei remains the same. 
And hence this shows that the two extra neutrons in $^{11} Li$ 
do not disturb the proton distribution of the core.

In addition, more significantly, the wave function which works for these 
halo nuclei is where the core and 
the halo neutrons decouple so that [7,8]

\begin{equation}
{{\Phi}_{halo nucleus} } =  
{{\Psi}_{core}} \bigotimes {{\Psi}_{2n}}
\end{equation}

There are other two neutrons halo nuclei known, eg.  
${^{14}_{4} Be_{10}}$ and
${^{17}_{5} B_{12}}$ etc. 
In fact there are also single neutron halo nuclei as well, eg 
${^{11}_{4} Be_{7}}$ and ${^{19}_{6} C_{13}}$.  So it is doubtful if
pairing has anything to do with the existence of halo nuclei.
In fact, through the determination of the momenta 
distribution of the halo neutrons, it 
has been found that they remain significantly far away from each other
in the halo [7].

The fact that the radii of neutron halo nuclei is so very large and this 
coupled with the fact that the wave function decouples as given above, 
is indicative of some "new physics". 
This point is consolidated by further studies of the halo 
nuclei. Below we discuss two more sets of empirical information 
arising from the study of halo and other exotic nuclei, which further 
strengthen the belief that this indeed points to some "new physics". 
This phrase "new physics" just means that we have to go 
beyond our understanding of nuclear physics based on conventional ideas.

In nuclear fusion process, the overlap between the two participating 
nuclei 
is important. Hence based on this picture of fusion in conventional 
nuclear physics, 
one would expect significant enhancement of the probability
of nuclear fusion at low energies if halo nuclei are involved.
A precision experiment to detect this effect was performed recently
by Raabe et al [9]. They studied the reaction of the halo nucleus
${^{6}_{2} He_{4}}$ on ${^{238}_{92} U_{146}}$ target at energies near 
the fusion barrier. Surprisingly, they found no such enhancement
which clearly contradicted the logic for halo nuclei based on 
conventional nuclear physics. This shows that the behaviour of 
halo nuclei is different from the standard nuclei on the basis of which 
all our understanding of nuclear physics is based upon.

Clearly the  two extra nucleons  in ${^{6}_{2} He_{4}}$ 
behave differently with respect to the other four 
nucleons in  the core nucleus ${^{4}_{2} He_{2}}$.
This is consistent with the total wave function of 
${^{6}_{2} He_{4}}$ being a simple product wave function as 
given above

\begin{equation}
{\Phi}_{{^{6}_{2} He_{4}} } =  
{\Psi}_{{^{4}_{2} He_{2}}} \bigotimes {{\Psi}_ {2n}}
\end{equation}

The symmetry inherent in such a product is that one does not 
antisymmetrize the two neutrons in the halo with respect to the
other neutrons in the core. Clearly this is not permitted for 
a standard single composite nucleus in conventional nuclear physics.
But this is consistent with all the other properties of halo nuclei.

And this is also consistent with the results of the fusion reaction
study discussed above [9]. What appears to be happening there, was that 
the large fission yields that they had obtained, did not result from the 
fusion with  ${^{6}_{2} He_{4}}$ but from neutron transfer. That is,
the last two neutrons in ${^{6}_{2} He_{4}}$ are first transferred to the 
target. This enriched target then reacts with the left over core of the 
halo nucleus. This is indeed a new phenomenon indicating the 
uniqueness of the halo nuclei.

Another new puzzling aspect arising from the study of exotic neutron rich 
nuclei is that of the sudden changing of magic numbers.
Magic numbers N,Z = 2,8,20,28,50,82 ... have been the corner stones of
conventional nuclear physics. 
However it has to be admitted that, though phenomenologically these had 
been incorporated in the shell model in nuclear physics, we really had no
basic understanding of how these magic numbers arose.
Studies of neutron rich nuclei have been 
indicating clearly that these conventional magic numbers are holy cows no 
more. Several aberrant magic numbers have been forcing themselves upon 
physicists. Studies of ${^{12}_{4} Be_{8}}$
and ${^{32}_{12} Mg_{20}}$ showed that N = 8 and 20 were magic no more for 
these nuclei [10,11].
The nucleus ${^{28}_{8} O_{20}}$ which should have been particularly 
stable as per the conventional nuclear shell model, was found to be not 
even bound [8]. New magic numbers like N=14,16 and 32 have been 
discovered. In fact N=14 in ${^{42}_{14} Si_{28}}$
was shown to be magic [12]. Clearly all these new magic numbers
are indicative of some "new physics" in nuclear physics [13].

One has to ask as to how halo is created, why does it behave so strangely
that there is no fusion enhancement due to its large size and also as to 
what leads to the changing of magic numbers. Here I shall try to provide 
some answers.

Going through the binding energy systematics of neutron rich nuclei one
notices that as the number of 
$ \alpha $'s increases along with the neutrons, each $ ^{4}He $ + 2n pair
tends to behave like a cluster of two 
$ ^{3}_{1}H_{2} $ nuclei. Remember that though 
$ ^{3}_{1}H_{2} $ is somewhat less strongly bound (ie. 8.48 MeV ) it is
still very compact (ie. 1.7 fm ), almost as compact as $ ^{4}He $ (1.674
fm). In addition it too has a hole at the centre. Hence $ ^{3}H $ is also
tennis-ball like nucleus. This splitting tendency of neutron rich nuclei
becomes more marked as there are fewer and fewer of $ ^{4}He $ nuclei left
intact by the addition of 2n. Hence $ ^{7}Li $ which is 
$ ^{4}He + ^{3}H $ with 2n becomes $ ^{9}Li $ which can be treated as made
up of $ 3 ~^{3}H $ clusters and should have hole at the centre. Similarly
$ ^{12}Be $ consists of 4 $ ^{3}_{1}H_{2} $ 
clusters and $ ^{15}B $ of 5 $ ^{3}_{1}H_{2} $  clusters etc.
Other evidences like the actual decrease of radius as one goes from
$ ^{11}Be $ to $ ^{12}Be $ supports the view that it ( ie $ ^{12}Be $ )
must be made up of four compact clusters of $ ^{3}H $.

Just as several light N=Z nuclei with A=4n, n=1,2,3,4 ... can be treated
as made up of n  $ \alpha $ clusters, in Table 1 we show several neutron
rich nuclei which can be treated as made up of n
$ ^{3}_{1}H_{2} $ clusters. We can write the binding energy of these
nuclei as 

\begin{equation}
E_{b} = 8.48 n + Cm 
\end{equation}
 
  where n 
$ ^{3}_{1}H_{2} $ clusters form m bonds and where C is the inter-triton
bond energy. We have assumed  the same geometric structure of clusters in
these nuclei as for $ \alpha $ clusters of A = 4n nuclei as given above.
All the bond numbers arise due to
these configurations. We notice from Table 1 that this holds good and 
and that the inter-triton cluster bond energy is approximately 5.4 MeV.
We notice that this value seems to work for even heavier neutron rich
nuclei. For example for 
$ ^{42}Si $ the inter-triton cluster energy is still 5.4 MeV. Notice that
the geometry of these cluster structures of $ ^{3}H $ becomes more complex
as the number increases but nevertheless, it holds well.

\vskip 7.0 cm

\begin{table}
\centerline {\bf Table 1} 
\centerline{ Inter-triton cluster bond energies of neutron rich nuclei }
\vskip 0.2 in
\begin{center}
\begin{tabular}{|c|c|c|c|c|}
\hline
Nucleus & n & m & $ E_{B} - 8.48n (MeV) $ & C(MeV) \\
\hline
$ ^{9}Li $ &  3 & 3 & 19.90 & 6.63 \\
\hline
$ ^{12}Be $ & 4 & 6 & 34.73 & 5.79   \\ 
\hline
$ ^{15}B $ & 5 & 9 & 45.79 & 5.09 \\
\hline
$ ^{18}C $ & 6 & 12 & 64.78 & 5.40 \\
\hline
$ ^{21}N $ & 7 & 15 & 79.43 & 5.29 \\
\hline

\end{tabular}
\end{center}
\end{table}

 The point is that these neutron rich nuclei, made up of n number of
tritons, each of which is tennis-ball like and compact, 
should itself be compact as well. 
These too would develop tennis-ball like property. This is
because the surface is itself made up of tennis-ball like clusters. 
Hence as there are no more $ ^{4}He $ clusters to break when
more neutrons are added to this ball of triton clusters, these 
extra neutrons will ricochet on the surface.
Hence we expect that one or two neutrons outside these compact clusters
would behave like neutron halos. Therefore $ ^{11}Li $ with $^{9}Li + 2n $
should be two neutron halo nuclei - which it is. So should 
$ ^{14}Be $ be. It turns out that internal dynamics of $ ^{11}Be $
is such that it is a cluster of $ \alpha - t - t $ ( which also has to do
with $ ^{9}Li $ having a good 3 $ \alpha $ cluster)
with one extra neutron halo around it. Next $ ^{17}B, ^{19}C, ^{20}C $
would be neutron halo nuclei and so on.

Hence, all light neutron rich nuclei $ _{Z}^{3Z}A_{2Z} $ are made up of Z 
$ ^{3}_{1}H_{2} $ clusters. Due to hidden colour considerations
arising from quark effects [13,14], all these should have holes at the 
centre. This would lead to tennis-ball like property of these nuclei.
One or two (or more) extra neutrons added to these core nuclei 
would ricochet on the surface of the core nucleus and
form halos around it. Practically all known and well-studied neutron halo
nuclei fit into this pattern. Also this
makes unambiguous predictions about which nuclei should be neutron halo
nuclei and for what reason. 
The proton halo nuclei can also be understood in the same
manner. Here another nucleus with a hole at the centre 
$ ^{3}_{2}He_{1} $ (binding energy 7.7 MeV, size 1.88 fm) would play a
significant role.
The success of this model here gives us confidence in the new picture 
proposed.

Note that 'n' and 'p' are members of an isospin doublet. These combine 
together to give a bound triplet state (S=1, T=0), that is deuteron with a 
binding energy of 2.2 MeV. It has no excited states. The singlet state 
(S=0, T=1) is unbound by 64 keV. This being isospin partner of (n-n) (S=0, 
T=1) and (p-p) (S=0,T=1), all these are unbound. 
Now it turns out that 'h' and 't' are also members of a good isospin 1/2. 
Note that though both 'n' and 'p' are composites of three quarks 
these still act as elementary particles as far as low-energy 
excitations of nuclear physics are concerned. Only at relatively higher 
energies does the compositeness of 'n' and 'p' manifest itself. Similarly
the binding energies of ${^3}He$ and ${^3}H$ are 7.72 MeV and 8.48 MeV 
respectively and also these two have no excited states. Hence for low-energy 
excitations, of a few MeV, we may consider these as elementary. Their 
compositeness would be manifested at higher excitation energies. That is 
we treat 'h' and 't' as elementary isospin 1/2 entities here. This is 
similar to the two nucleon case. Hence, we would expect for (h-t) the 
triplet (S=1,T=0) to be bound and singlet (S=0,T=1) to be unbound. Also, its 
isospin partners (h-h) (S=0, T=1) and (t-t) (S=0, T=1) would be unbound too. 
Herein triton ("t")  ${^{3}_{1} H_{2}}$
helion ("h")  ${^{3}_{2} He_{1}}$ are treated as
fundamental representations of this new symmetry group  
called "nusospin" $SU(2)_{\cal A}$. 
Even the isospin symmetry between neutron and proton is broken.
So we expect that this new symmetry should be broken as well.

To see how this new symmetry is likely to manifest itself
we look at other symmetry groups which are known to be broken,
So in quark model we know that progressively the flavour symmetry
groups SU(2), SU(3), SU(4) etc for more number of quark flavours
are broken more and more strongly. In fact SU(5) for five quark flavours
u,d,s,c and b quarks is very strongly broken. However it still manifests 
itself in particle physics. Its  most clear manifestation is 
in terms of 
representations of all the particles built up of any of these 5 quarks.
Hence, howsoever badly it may be broken, 
the physical existence of objects which correspond to the 
irreducible representation of a particular group is what actually
determines the relevance of a particular group in physics.
So also in the case of our new symmetry in nuclei - 
$SU(2)_{\cal A}$ this may be minimum expectation
as well. Hence as already suggested, the particle representation 
of nuclei of the form ${^{3Z}_Z} A _{2Z}$ nuclei 
would be that of Z number of tritons as
per the "nusospin" group..

So let us ask as to what this new "nusospin" symmetry 
$SU(2)_{\cal A}$ has to tell us about
new magicities. Clearly the fact that  
${^{3Z}_Z} A _{2Z}$ nuclei 
are made up of Z number of tritons leads to new stability for them.

We plot
$S_{1n}$ as a function of $Z$ for a particular $N$
rather than plotting it as a function of $N$,
as is normally done [11,12]. 
It appears that what we are plotting brings out the relevance of
the (h,t) degree of freedom of the new group nusospin more clearly.
Fig 1 is $S_{1n}$ for fixed N=
 4, 6, 8, 10, 11, 12, 16, 20, 22 and 24 plotted as a function of Z.

The reader's attention is drawn to the extra-ordinary stability manifested 
by the plotted data for the proton and neutron pairs (Z,N): 
(6,12), (8,16), (10,20), (11,22) and (12,24). 
Note that the stability at these pair
of numbers is sometimes as prominent as that at the N=Z pair. In fact 
the Z,N pair (10,20) stands out as the best example of this.  
Hence it is clear that the separation energy data very clearly shows 
that there are new magicities present in the neutron rich sector for the 
pair (Z,N) where N=2Z.
For more plots of this kind which show stability of (Z,2Z) nuclei,
see [13].

What is the significance of this extraordinary stability or magicity for
the nuclei ${^{3Z}_Z} A _{2Z}$? We already know that for the even-even Z=N 
cases it is the significance of $\alpha$ clustering for the ground state 
of these nuclei which explains this extra stability.
Quite clearly the only way we can explain the extra magicity for these
N=2Z nuclei is by invoking the significance of triton clustering
in the ground state of these neutron rich nuclei. 
Thus ${^{30}_{10} Ne _{20}}$ has significant mixture of the 
configuration 10 $^{3}_{1} H_{2}$ in the ground state.
It is these tritonic clusters which give the extra stability to these 
nuclei thereby providing us with these unique new
sets of magic numbers.

We continue to plot the separation energies a little differently here.  
We plot $S_{1n}$ and $S_{2n}$ as a function of $Z$ for a particular $N$.
The same with  $S_{1p}$ and $S_{2p}$ as a function of N for a particular
Z. We do a systematic study of these plots for all the data 
available in literature at present. 
We find that these brings out certain very 
interesting generic features which as we shall find clearly 
indicate strong evidences of   
triton ("t")  ${^{3}_{1} H_{2}}$
and helion ("h")  ${^{3}_{2} He_{1}}$ 
clustering in nuclei

We study the whole range of data set available. 
For the sake of brevity, here we 
show two representative plots. These are 
one and two neutron separation energy as a function of Z 
for N = 29 and 30 as plotted in Fig 2 and 3 below.
All the plots of 
$S_{1n}$ and $S_{2n}$ as a function of $Z$ for a particular $N$
show similar features.
The same with  $S_{1p}$ and $S_{2p}$ as a function of N for a particular
Z. We do a systematic study of these plots for all the data.
The features which we shall point out here are there in all the 
other plots. In fact we shall study here only those features which 
are generic of all such plots. 

Certain common features which stand out are as follows
(The statements below are made in the context of
the plot
$S_{1n}$ and $S_{2n}$ as a function of $Z$ for a particular $N$):

\vskip .5 cm

{\bf A}. {\it For all even-even N=Z nuclei there is always a 
pronounced larger separation energy required with respect to the 
lower adjoining nuclei plotted}.

\vskip .5 cm

{\bf B}. {\it For all odd-odd N=Z nuclei there is always a 
pronounced larger separation energy required with respect to the 
lower adjoining nuclei plotted}.

\vskip .5 cm

{\bf C}. {\it For case {\bf A} when Z number is changed by one 
unit, the separation energy hardly changes (sometimes not at all). 
But when this number is changed by two units, another 
pronounced peak occurs}.

\vskip .5 cm 

{\bf D}. {\it For case {\bf B} when Z  number 
is changed by one unit, the 
separation energy hardly changes (sometimes not at all). But when 
this number is changed by two units, another pronounced peak occurs.
So there are peaks for odd-odd N-Z nuclei ( and not for 
even-even cases )}.

\vskip .5 cm 

Here, as we are pulling one or two neutrons as a function of Z,
the above effects cannot be the result of
identical nucleon pairings, What these plots are telling us is as
to what happens to last one or two neutron bindings in a nucleus 
as proton number changes.

Clearly the peaks as indicated in {\bf A} above are due to the fact 
that the last one or two neutrons must have come from a stable 
alpha cluster. This is consolidated by the fact that another extra Z 
does not make a difference to the separation energy.

To understand observation {\bf B} 
above - note that here the
peaks are there for ALL odd-odd N=Z nuclei. This is an amazing
fact. These odd-odd nuclei are more stable or "magic" with respect 
to the adjoining odd-even or even-odd nuclei.
Pairing of identical nucleons cannot explain this generic 
feature. Neither can alpha clustering do so.
Obviously it is the formation of triton-helion 'h-t' pair
which can only explain this extraordinary effect.

What is the significance of this extraordinary stability or magicity for
all the nuclei ${^{3Z}_Z} A _{2Z}$? 
Quite clearly the only way we can explain the extra magicity for these
N=2Z nuclei is by invoking the significance of triton clustering
in the ground state of these neutron rich nuclei. 
So the nucleus ${^{3Z}_Z} A _{2Z}$ is made up of Z number of 
triton clusters, as we showed earlier.

To understand this unique feature, the new
symmetry "nusospin" symmetry becomes relevant.
It is clear that it is tritons which explain the stability of 
neutron rich nuclei and it it is pair of 'h-t' clusters which 
explain the stability of odd-odd N=Z nuclei in the separation 
energy as plotted above.

Now we can explain the observation {\bf C} above. Clearly for 
even-even N=Z nuclei it is one ( or more ) alpha clusters which 
explain the data. Hence one extra Z does not affect the 
separation energy. But two extra Z will tend to make a
pair of helions ( akin to the two tritons for neutron rich 
case above ).
These two 'h-h' will make for the extra stability for the adjoining 
even-even nuclei (and so on). 
So also can the observation {\bf D} be 
understood as the extra 2Z will create an extra helion to attach 
to the already existing 'h-t' pair to make for extra stability 
of this adjoining odd-odd nuclei. Other peaks in the above plots 
can be similarly explained as due to alpha or triton and helion 
clusters. Clearly the new nusospin group provides the (h,t) degrees of 
freedom to explain the above data. 

As we have found the nusospin group to be useful in certain physical
situations, the relevant enlargened group in nuclear physics should be

${{SU(2)}_{I}}  \bigotimes  {{SU(2)}_{\cal A}}$

ie a product of the isospin and the nusospin groups.
Hence both the (p,n) and (h,t) are relevant degrees of freedom for 
nuclei. 

This helps us in resolving a puzzle in the structure of 
${^{4}_{2} He_{2}}$ nucleus. It is now well known [17], that contrary to 
expectations, the ground state of 
${^{4}_{2} He_{2}}$ contains very little of deutron-deutron
configuration; and the same is actually built upon h-n and t-p
configurations [17]. It is a puzzle as to how come the first excited 
state of this even-even nucleus is 
another $ 0^{+} $ with T=0 state ( the same as the 
ground state ) at a high value of 20.2 MeV.
However this finds a natural explanation in our model of the 
product group ${{SU(2)}_{I}}  \bigotimes  {{SU(2)}_{\cal A}}$
The wave functions of the ground state and the first excited state
of ${^{4}_{2} He_{2}}$ in our model is naturally given as

\newpage

\begin{equation}
{{\Phi}_{gs}} =  
{{ [ {{\Psi}_{h}}  \bigotimes {{\Psi}_{n}} - {{\Psi}_{t}} \bigotimes 
{{\Psi}_{p}} ] } 
\over {\sqrt {2}}}
\end{equation}

\begin{equation}
{{\Phi}_{20.2}}  =
{{ [ {{\Psi}_{h}}  \bigotimes {{\Psi}_{n}} + {{\Psi}_{t}} \bigotimes 
{{\Psi}_{p}} ] }
\over {\sqrt {2}}}
\end{equation}

Clearly this enlargened group 
${{SU(2)}_{I}}  \bigotimes  {{SU(2)}_{\cal A}}$
should help us in improving the understanding of nuclear phenomenon.

So in summary, the field of nuclear physics today is an active and a very
promising branch of physics. The new results are forcing us to rethink
some of the fundamentals of conventional nuclear physics. 
Obviously what has been working ( and hence phenomenologically correct )
for N=Z and nearby nuclei cannot be wrong. 
But we have to 
extend our understanding so as to incorporate the new puzzling data 
arising mainly from the study of exotic nuclei. The N=Z and nearby cases 
should be considered as special case of a more general framework which 
should account for the exotics as well. We have tried to present a 
picture which tries to tackle some of these issues.
Coming up with a comprehensive picture is the exciting 
challenge of nuclear physics today.

\newpage

\begin{figure}
\caption{One neutron separation energy as a function of 
proton number Z for different number of fixed neutrons indicated}
\epsfclipon
\epsfxsize=0.99\textwidth
\epsfbox{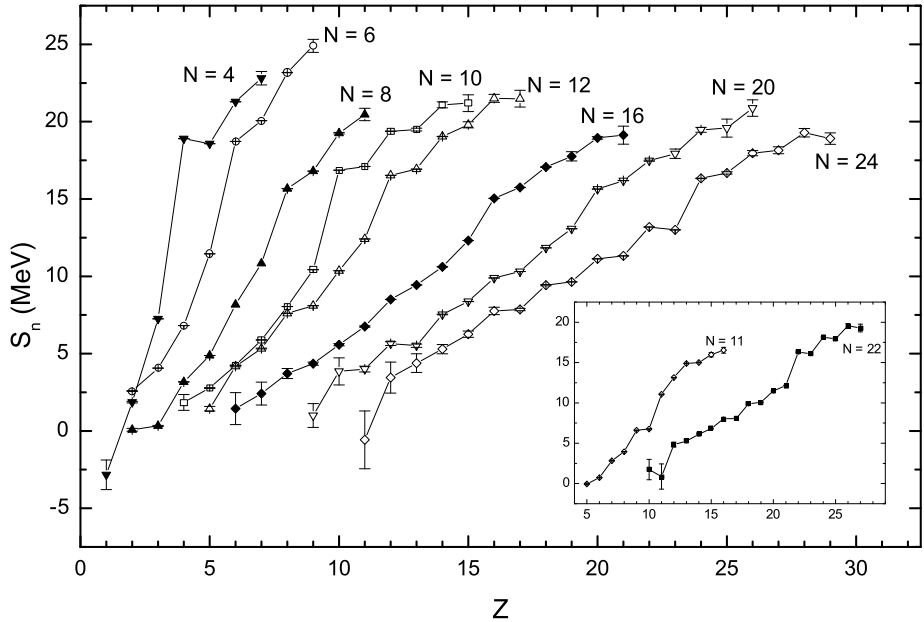}
\end{figure}

\newpage

\begin{figure}
\caption{One and two neutron separation energy as a function of 
proton number Z for fixed N=29 neutrons}
\epsfclipon
\epsfxsize=0.99\textwidth
\epsfbox{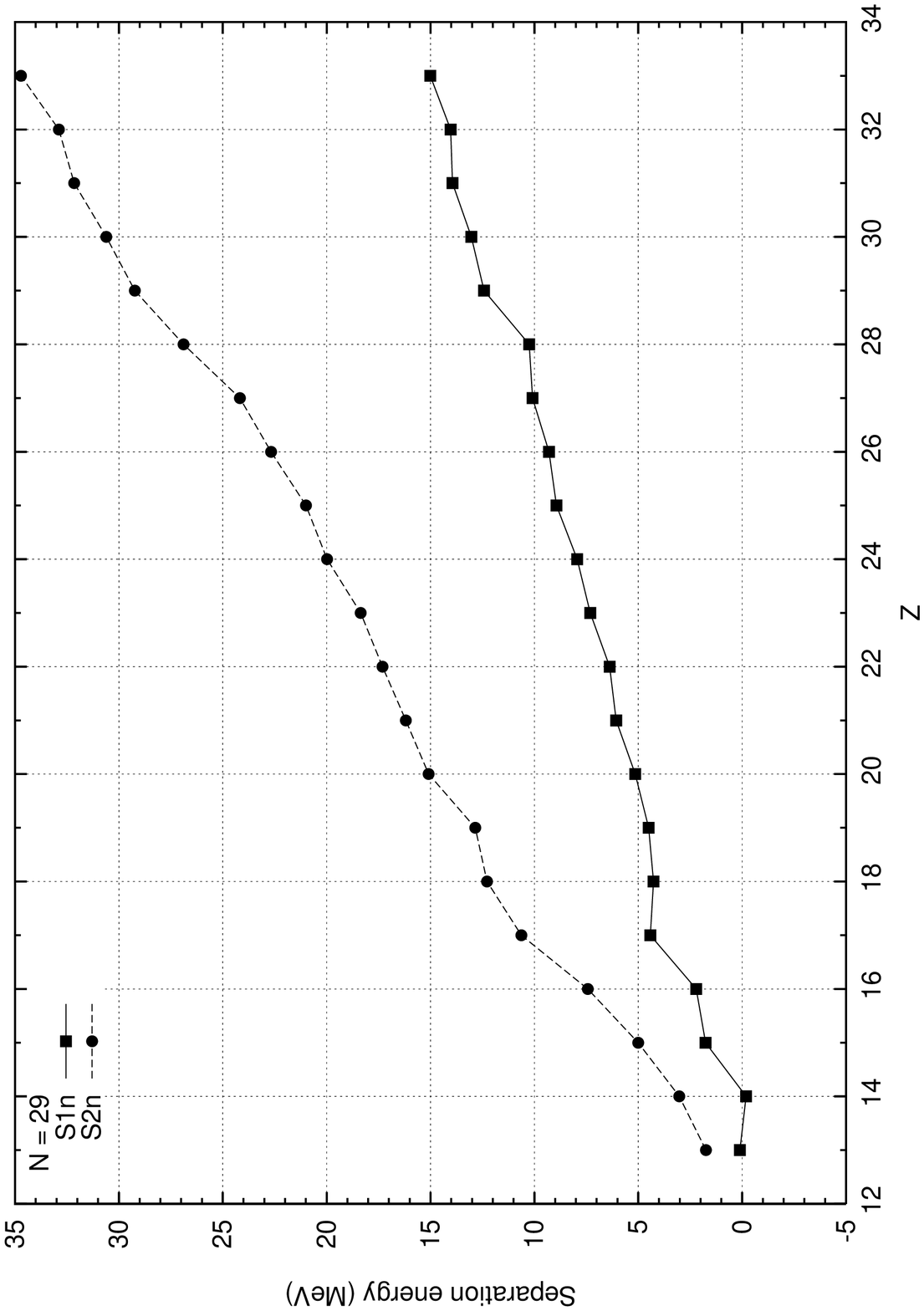}
\end{figure}

\newpage

\begin{figure}
\caption{One and two neutron separation energy as a function of 
proton number Z 
for fixed N=30 neutrons}
\epsfclipon
\epsfxsize=0.99\textwidth
\epsfbox{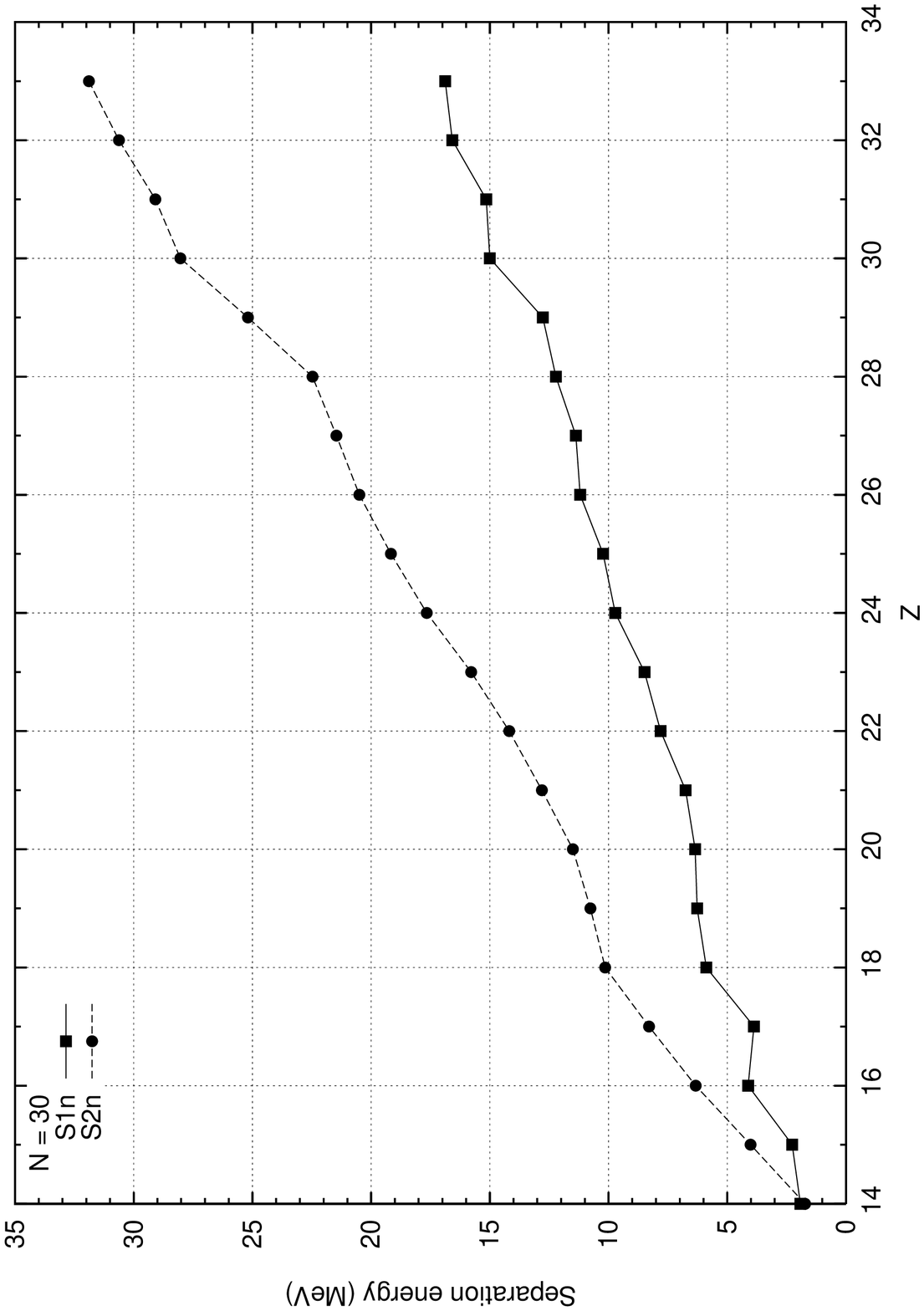}
\end{figure}

\newpage

\vskip .3 cm

\begin{center}
{\bf REFERENCES }
\end{center}
\vspace{.4in}

1. A. Richter, Nucl. Phys. {\bf A 533} (1993) 417c 

2. RHIC Report: Conceptual Design of Relativistic Heavy Ion Collider, 
   BNL Report 52195, Brookhaven National laboratory
  
3. B. Mueller, "The physics of quark gluon plasma", (Lecture Notes in 
   Physics Vol 225), Springer Verlag, Berlin 1985

4. A. Abbas, L. Paria and S. Abbas, Eur. Phys. J. {\bf C 14} (2000) 695

5. P. J. Twin et al, Phys. Rev. Lett. {\bf 57} (1986) 811

6. W. Gelletly, Contemp. Phys. {\bf 42} (2001) 285

7. I. Tanihata, J. Phys. {\bf G 22} (1996) 157

8. I. Tanihata, Nucl. Phys. {\bf A 685} (2001) 811

9. R. Raabe et al, Nature {\bf 43} (2004) 823

10. Z. Dlouhy, D. Baiborodin, J. Mrazek and G.
    Thiamova, Nucl. Phys. {\bf A 722} (2003) 36c

11. M. Thoennessen, T. Baumann, J. Enders, N. H. Frank, P. Heckman, 
    J. P. Seitz and E. Tryggestad, Nucl. Phys. {\bf 26} (2003) 61c

12. J. Fridmann et al, Nature {\bf 435} (2005) 922

13. A. Abbas, Mod. Phys. Lett. {\bf A 20} (2005) 2553

14. A. Abbas, Phys. Lett. {\bf B 167} (1986) 150

15. A. Abbas, Mod. Phys. Lett. {\bf A 16} (2001) 755

16. G. Audi, A. H. Wapstra and C. Thibault, Nucl. Phys. {\bf A 729}
    (2003) 337

17. K. Langanke, Adv. Nucl. Phys. {\bf Vol 21} (1994) 85

\end{document}